\documentstyle[12pt,epsfig]{article}
\begin{document}

\begin{titlepage}
\begin{flushright}
hep-th/9802149 \\
February 14, 1998\\
\end{flushright}

\vskip 2.cm

\begin{center}
{\Large\bf A New Variant of the Casimir Effect and Its Exact Evaluation.}
\vskip 2.cm

{\large Oded Kenneth and Shmuel Nussinov}

\vskip 0.5cm

\vskip 1cm
{\it School of Physics and Astronomy,\\ Raymond and Beverly Sackler
Faculty of Exact Science,\\ Tel-Aviv University,\\ Tel-Aviv 69978, Israel}
\end{center}
\vskip 3cm

\begin{abstract}

{\baselineskip=24pt A new version of the Casimir effect where the two
plates conduct in specific, different, directions is considered.  By
direct functional integration the evaluation of the Casimir energy as
a function of the angle between the conduction directions is reduced
to quadratures.  Other applications of the method and a magnetic
Casimir variant are mentioned.}

\end{abstract}
\vfill
\end{titlepage}

The Casimir force per unit area [1] 
\begin{equation}
\frac{F_{\rm cas}(a)}{A} = -\frac{\pi^2}{240} \frac{\hbar c}
{a^4}
\end{equation} 
attracts two parallel conducting plates at a distance $a$ apart
in vacuum.  Its independence of atomic and QED parameters reflects
the perfect conductor idealization where all details are subsumed into
boundary conditions
$E_x = 0,~~E_y = 0~\hbox{at}~z=0~\hbox{or}~z = a$. 
These, in turn, quantize
the $z$ component of 
the wave number vector for modes in the region between the plates,
$k_z = \frac{n\pi}{a}$.  The problem then reduces to evaluating the
change in vacuum energy of all the transverse modes inside this region:
\begin{equation}
\frac{E_{\rm cas}}{A} =
\int dk_xdk_y
\left\{\sum_n\sqrt{k_x^2+k_y^2+\left(\frac{n\pi}{a}\right)^2} ~
-\frac{a}{\pi} \int dk_z \sqrt{k_x^2+k_y^2+k_z^2}\right\} 
\end{equation}
A careful regularization of this formally divergent expression yields
[2] \break $E_{\rm cas}(a)/A = -\pi^2\hbar c/(720 a^3)$ and $F_{\rm
cas} = -\frac{d}{da} E_{\rm cas}(a)$.

The tiny Casimir forces  are elusive.  Past efforts [3] 
verified Eq. (1) rather roughly and only recently a 5\% precision
experiment was done [4].

In this paper we focus mainly on
a new variant of the Casimir effect: each of the two plates
conduct in a specific direction: $\hat e_1$ for plate one and $\hat e_2$
on plate two so that only the components $\vec E\cdot\hat e_1$ and
$\vec E\cdot\hat e_2$ need to vanish on plates I and II respectively, and
$\hat e_1\cdot\hat e_2\equiv\cos\beta$ is arbitrary.
  
The two polarizations contribute equally to $F_{\rm cas}$.  By
the above ``twist" these polarizations can generate a controlled
$W_{\rm cas}^{(\beta)},F_{\rm cas}^{(\beta)}$ and a Casimir torque,
$\tau_{\rm cas}^{(\beta)}$.

The Casimir force can be derived also [5] by evaluating the pressure imbalance
due to reflection of ``vacuum modes" off the outside surface of the plate and
of the (fewer) internal modes off the inside surfaces.
  
If the two plates are replaced by arrays of  
only vertical (or horizontal) conducting wires (mimicking the
anisotropic conductivities) then 
only the $\hat e_y(\hat e_x)$
polarized modes will be 
reflected suggesting that we have half the Casimir force.
We also expect further reduction [6] of the Casimir force as $\beta$, the
angle between the directions of the two sets of wires (or the directions
of conductivity in the two plates) increases from $\beta = 0$ to $\beta = 90$.

We will next present an exact evaluation of $W_{\rm cas}(\beta,a)$ for
general $\beta$'s using an altogether different
method [7].
We add to the free electromagnetic action $S_{\rm em} = -\frac{1}{4} \int
F^2d^4x$ a Lagrange multiplier term $\int J\cdot Ad^3x$ (where the last
integral is only over the area-time of the plates).  The functional integration
over $DJ$ ensures the vanishing of the transverse electric field $E_T$ over
the plates.  Since we want the $A_\mu$ field to vanish only up to gauge
transformations we integrate only over conserved currents $J$.  Also since 
only transverse components of $E$ have to vanish we allow $J$ to have
components only in the three-dimensional area-time of the plates i.e.
$J = (J_x,J_y,J_t)$.  Thus we will write for the partition function
\begin{equation}
Z = \int DADJ \exp \left(-i\int d^4x\frac{1}{4} F^2 +
i\int d^3xA\cdot J\right)
\end{equation}
The only difference in the case of interest, where each plate conducts in
a specific direction, is that only one component of $E_T$ has to vanish on
each plate.  This can be achieved by further restricting the allowed
currents $J$.  Hence Eq. (6) and its consequences remain correct as long
as we remember to interpret $\int DJ$ differently.  Changing the order of
integration and doing first the standard $DA$ integration yields:
\begin{equation}
Z = \int DJ\exp\left(-\frac{i}{2}\int J_\mu(x)\Delta_F(x-y)
J^\mu(y)d^3xd^3y\right)
\end{equation}
where $\Delta_F(x-y) = \frac{1}{4\pi^2i}\frac{1}{(x-y)^2-i\epsilon}$ is
the massless Feynman propagator [8] 
(constant's coefficients such as the $1/4\pi^2$ 
above  contribute only an overall multiplicative
factor or an additive term to the energy and will
be discarded henceforth). Denoting the currents on the first and second plates by
$J_1$ and $J_2$, we have the more explicit expression for $Z$:
\begin{equation}
\int DJ \cdot \exp -i\int d^3xd^3y
\left(\frac{J_1(x)\cdot J_1(y)+J_2(x)\cdot J_2(y)}{(x-y)^2-i\epsilon}
+\frac{2J_1(x)\cdot J_2(y)}{(x-y)^2-a^2-i\epsilon}\right)
\nonumber
\end{equation}
or after a Wick rotation
\begin{equation}
\int DJ \cdot \exp-\int d^3xd^3y
\left(\frac{\vec J_1(x)\cdot \vec J_1(y)+\vec J_2(x)\cdot\vec J_2(y)}
{(x-y)^2} +
\frac{2\vec J_1(x)\cdot\vec J_2(y)}{(x-y)^2+a^2}\right)
\nonumber
\end{equation}
where we think of $\vec J_{1,2}$ as of ordinary 3-vectors in ordinary
3-dimensional Euclidean space (although it is actually spanned by
$x,y,t$).  Fourier transforming in $\vec x=(x,y,t)$
this becomes:
\begin{equation}
\int DJ(\vec k) 
\exp~-\int d^3k
\left(\frac{\vec J_1(\vec k)\cdot\vec J_1(-\vec k)+\vec J_2(\vec k)\cdot
\vec J_2(-\vec k)}{k} + \frac{2\vec J_1(\vec k)\cdot\vec J_2(-\vec k)}{k}
e^{-ka}\right) \nonumber
\end{equation}
where $\vec k = (k_x,k_y,k_t),~k=|\vec k|$ and we used translation invariance.
In the usual case of two conducting plates both $\vec J_1(k)$ and
$\vec J_2(k)$ have two 
transverse degrees of freedom 
fixed by the
current conservation condition:
$\vec k\cdot\vec J = 0$.  In the case of specific conduction directions,
$\vec J_1(k)$, and likewise $\vec J_2(k)$, have only one allowed non-zero
component determined by current conservation and by the demand that its
spatial part $(J_x,J_y)$ is along the direction of conduction.  Let us
denote the cosine of the angle between the directions of
$\vec J_1(\vec k)$ and $\vec J_2(\vec k)$ by $\alpha(\vec k)$ (with
$\vec J_1,\vec J_2$ an ordinary Euclidean vector).  Then we can write for $Z$:
\begin{equation}
\int DJ(\vec k) 
\exp~-\int d^3k
\left(\frac{J_1(\vec k)J_1(-\vec k)+J_2(\vec k)J_2(-\vec k)}{k} +
\frac{2J_1(\vec k)J_2(-\vec k)\alpha(k)}{k} e^{-ka} \right)
\nonumber
\end{equation}
where the $J_i(\vec k) \vec J_i$
are scalars and the reality of $J(x)$ implies:
\begin{equation}
[J_i(\vec k)]^* = J_i(-\vec k)
\end{equation}

Since the action is quadratic, $Z$ is given by the corresponding determinant
which is just the product of the two-dimensional determinants corresponding
to the various value of $\vec k$.  Hence
\begin{equation}
Z = \prod_{\vec k} \det\left(
\begin{array}{cc}
\frac{1}{k} & \frac{\alpha\hat k)}{k} e^{-ka} \\
\frac{\alpha(\hat k)}{k}e^{-ka} & \frac{1}{k} 
\end{array}
\right)^{-1/2}
\end{equation}
\begin{eqnarray}
\ln Z &=& -\frac{1}{2} \sum_{\vec k}\ln \det(\ldots) \nonumber \\
&=& -\frac{1}{2} AT\int \frac{d^3k}{(2\pi)^3} \ln
\left|
\begin{array}{cc}
\frac{1}{k} & \frac{\alpha(k)}{k} e^{-ka} \\
\frac{\alpha(k)}{k} e^{-ka} & \frac{1}{k}
\end{array}
\right| \nonumber \\
&=& -\frac{1}{2} AT\int \frac{d^3k}{(2\pi)^3} \ln
\left(\frac{1-\alpha^2e^{-2ka}}{k^2}\right) \nonumber \\
&=& -\frac{1}{2} AT\int \frac{d^3k}{(2\pi)^3}
\ln(1-\alpha(\hat k)^2e^{-2ka})+\hbox{const.}
\end{eqnarray}
where the area-time $AT$ came from density of states factor [9].  
It corresponds
to having
\begin{equation}
\sum_k \to V\int \frac{d^3k}{(2\pi)^3} \nonumber
\end{equation}
the usual quantization of continuous modes in a box of volume $V$.   Note that the last integral
in Eq. (11) is well defined and convergent.
To obtain it, we discarded the infinite
\begin{equation}
AT \int \frac{d^3k}{(2\pi)^3} \ln k^2 \nonumber
\end{equation}
term which does not depend on $a$ or the angle $\beta$ between the 
directions of conductivity in the two plates and hence does not contribute
to Casimir forces/torque [9].  Identifying $\ln Z = -ET$ we get finally:
\begin{equation}
\frac{E}{A} = \frac{1}{2} \int \frac{d^3k}{(2\pi)^3}
\ln(1-\alpha(\hat k)^2e^{-2ka})
\end{equation}
or using integration by parts (with respect to $k$ after separating
$\int d^3 k = \int d\Omega \int k^2dk$):
\begin{eqnarray}
\frac{E}{A} &=& -\int \frac{d^3k}{(2\pi)^3}~\frac{ka}{3}~
\frac{\alpha(\hat k)^2e^{-2ka}}{1-\alpha(\hat k)^2e^{2ka}} \nonumber \\
&=& -\frac{1}{48a^3}\int\frac{d^3k}{(2\pi)^3}~
\frac{k\alpha(\hat k)^2}{e^k-\alpha(\hat k)^2}
\end{eqnarray}
where in the last step we multiplied numerator and denominator by
$e^{2ka}$ and changed variables $2ka\to k$.

We next find an explicit expression for $\alpha(\hat k)$.  To this end,
let us denote by $\hat n_1,\hat n_2$ the two unit vectors in the planes of
plates and perpendicular to the direction of conduction in the first and
second plate respectively. 
The direction $\hat J_i$ is then determined by the
conditions $\hat k\cdot\vec J_i = \hat n_i\cdot\vec J_i = 0$.  Hence
\begin{equation}
\alpha=\cos(\vec J_1,\vec J_2) = \cos(\vec k\times\hat n_1,\vec k\times
\hat n_2) =
\frac{(\vec k\times\hat n_1)\cdot(\vec k\times\hat n_2)}
{|\vec k\times\hat n_1||\vec k\times n_2|} \nonumber
\end{equation}
Choosing $\hat n_1=(1,0,0),~\hat n_2=(\cos\beta,\sin\beta,0)$ and using
polar coordinates decomposition of $k$:  $\vec k = (k\sin\theta\cos\varphi,
k\sin\theta\sin\varphi,k\cos\theta)$ we find
\begin{equation}
\alpha^2 = \frac{[\cos\beta-\sin^2\theta\cos\varphi\cos(\varphi-\beta)]^2}
{(1-\sin^2\theta\cos^2\varphi)(1-\sin^2\theta\cos^2(\varphi-\beta))}
\end{equation}
For $\beta=0$ the integral can be calculated analytically
\begin{equation}
\frac{E_0}{A} = -\frac{\pi^2}{1440 a^3} \nonumber
\end{equation}
In this case $\alpha\equiv 1$
and Eq. (14) yields
\begin{eqnarray}
\frac{E}{A} &=& -\frac{1}{48a^3}\int\frac{4\pi}{(2\pi)^3}
\frac{k^3dk}{e^k-1} = -\frac{1}{96 a^3\pi^2} \sum_n \int^\infty_0
k^3e^{-nk}dk \nonumber \\
&=& -\frac{6\sum_n\frac{1}{n^4}}{96a^3\pi^2} =
\frac{-\frac{6\pi^4}{90}}{96a^3\pi^2} = -\frac{\pi^2}
{1440a^3}\nonumber
\end{eqnarray}
This is exactly half the usual Casimir energy, as expected since only one
polarization contributes providing a nice check of the calculation.

For general $\beta$
numerical integration of Eq. (15) yields
$E(a,\beta)/E(a,0)$ 
as plotted in Fig. 1.

\begin{figure}
\begin{center}
\epsfig{file=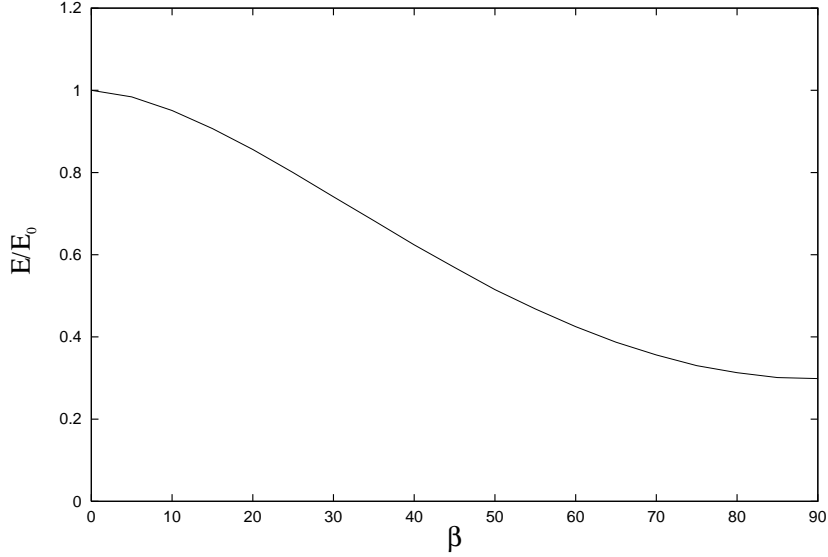,width=3in,angle=-90}
\caption{ The Casimir energy versus the angle between the directions
of conduction. }
\end{center}
\end{figure}

\vskip1.0cm

In addition to the $\beta$ dependent $F_{\rm cas} =
-\frac{\partial}{\partial a} E(a,\beta)$,
\begin{equation}
-\frac{\partial}{\partial\beta} E_{\rm cas}(a,\beta) =
\tau_{\rm cas}(a,\beta)
\end{equation}
is a Casimir torque tending to align the plates so that $\beta=0$,
i.e.  parallel conductivity directions and minimal energy are
achieved.  Since torques are easier to measure, $\tau_{\rm
cas}(a,\beta)$ could perhaps be tested to a better accuracy than
$F_{\rm cas}(a)$ or $F_{\rm cas}(a,\beta)$ --though extreme, global,
flatness will be required to avoid friction due to microscopic
roughness, as the circular plates rotate relative to each other.

We note that for non-circular plates, or when we have broad conducting
stripes, another aligning torque results from the preference to have
maximal adjacent plate area.  Clearly, this trivial geometrical effect
is not the subject of interest here.  The following comments are in
order:
\begin{description}
\item{(i)} The mode sum/integral in Eq. (2) involved in the usual derivation
is very different from that in the present derivation.  In particular, beyond
our discarding of the $a,\beta$ independent infinite part
in the last step of Eq. (11), no regularization is required here.  Also  
extending the standard method to derive our result for $\beta\not= 0$,
seems rather difficult.
\item{(ii)} The only relevant distance in this problem is the space-like
plate separation involved here.  This problem as well as any other static
Casimir calculation can therefore be done directly by the Euclidean path
integral with $Z = e^{-ET}$ and no Wick rotation is needed.
\item{(iii)} The present method was applied also to other geometries
(spherical, cylindrical, etc.)  The known Casimir energies were retrieved
though not with extra ease or rigor.
\item{(iv)} For the case of two magnetic/electric polarizable objects much
smaller than their separation the present formalism readily yields the
usual Casimir--Polder result for the potential between polarizable atoms [10].
\item{(v)} Finite temperature
can be easily incorporated
into the present approach, by replacing the $K_t$ integration in Eq. (14)
by a discrete sum.  We can also generalize to the case of $n \not= 3$
dimensions, by replacing $d^3k$ there by $d^nk$.
\item{(vi)} The present approach can be readily
generalized to time dependent boundary condition.  As noted also by
Golestanian and Kardar (see Ref. 7 above), this can result in ``Casimir
radiation".    
In particular we find 
that the lowest order (two-photon amplitude is simply given by
\begin{equation}
\langle 0|k_1\vec\epsilon_1,k_2\vec\epsilon_2\rangle =
\alpha w_1w_2(\epsilon_1\cdot\epsilon_2)\int d\tau \exp[i(k_1^\mu+k_2^\mu)
\cdot x_\mu(\tau)]
\end{equation}
with $\alpha$ the polarizability of
a neutral and small (relative to $\lambda$) object which
moves along a trajectory $x_\mu(\tau)$ with $\tau$ the proper time.
\end{description}
All of the points (iii)-(iv) will be elaborated in a longer paper by 
O. Kenneth.

Returning to the main theme of this Letter, we note that the naive argument
of Ref. (6) that $W_{\rm cas}(\beta = 90^\circ)$ should vanish [due to
the fact that for two arrays of orthogonal wires, the $\hat x$ polarized
modes are free to escape from the left, say, and the $\hat y$ polarized modes
from the right] fails.  As indicated in Fig. 1, this is definitely not the
case.  Indeed one simple approach views the ordinary attractive Casimir
effect as the attraction between patches of charges formed on one plate by
charge separation due to quantum fluctuations and patches of opposite sign,
``Image", charges induced on the other plate.  Clearly this mechanism can
operate, albeit with reduced strength, even if the two plates are made of
conducting stripes pointing in orthogonal directions.

This brings us to the final subject that we would like to mention, namely,
an analog interpretation of magnetic Casimir attraction between two
conducting rings.

Consider then two parallel conducting
rings of size $a$ and at a
distance $a$ apart.  The magnetic vacuum
fluctuations include closed
B field lines 
which link both rings.
These will induce, by Faraday's law, parallel
currents in the two rings.
Thus, regardless of the sign of the
B fluctuation and of the ensuing circulating current,
the resulting current
current forces will be attractive [12].

The fluctuations of interest are of scale $\lambda \approx a$ when the
above current - current forces on the various segments of the rings
add coherently corresponding to the net current flow in the rings
$R_1, R_2$.  If there is no net global flux change in the rings due to
the vacuum fluctuation there will be - in this approximation - no net
current and no net force.

How will this force be modified if the rings become
superconducting?  

Ideally,
the superconducting rings impose a new integral
constraint, namely that the total fluxes threading the various
superconducting rings must be
integer multiples of the flux quantum:
$\Phi = n\Phi _{0} = \frac{n\pi \hbar}{e}$.

This implies however a strong exponential
suppression.  Thus if we have a
fluctuation with roughly
constant $B$ on scale $a$:

\begin{equation}
\pi B a^{2} \approx n\Phi _{0} \approx \frac{n\hbar}{e}
\end{equation}
The action of such a configuration will therefore be:

\begin{equation}
A=  \int (cB)^{2} d^{3}x dt \approx \pi^{2} c^{2}
B^{2} a^{3} \frac{a}{c} = \pi^{2}c(Ba^{2})^{2} =
\frac{cn^{2}\hbar^{2}}{e^{2}}
\end{equation}

The exponential suppression 
$\exp\left[
-\frac{A}{\hbar}\right] 
\approx \exp\left[-\frac{n^{2}}{\alpha_{em}}\right]$
renders such fluctuation and the attendant
magnetic Casimir forces completely negligible.

The above considerations suggest
that if the Casimir force between
conducting rings is constantly monitored as
the temperature of the
system is lowered below the superconducting
critical temperature, then the
quenching of part of the magnetic Casimir force 
reduces the observed effect.
Hopefully, this amusing effect can eventually be observed,
but we will not elaborate here on the conditions necessary
for this.

\section*{Acknowledgment}
We would like to thank Z. Nussinov
for pointing us to us the recent quant-ph reference by Golestanian and
Kardar, and to Gilda Reyes for help with the preparation of the
manuscript.

\section*{References and Remarks}
\begin{description}
\item{[1]} H. B. G. Casimir, Proc. K. Ned Akad Wet. {\bf 51}, 793
(1948).
\item{[2]} See, e.g., C. Itzykson and J. B. Zuber, {\it Quantum
Field Theory}, McGraw Hill (1985), p. 137.
\item{[3]}M. J. Sparany, {\it Nature} {\bf 180}, 334 (1957) and
{\it Physica} {\bf 24}, 751 (1958).
\item{[4]} S. K. Lamoreaux, Phys. Rev. Lett. {\bf 78}, 5 (1997).
\item{[5]} P. W. Miloni, R. J. Cook, and M. E. Goggin, Phys. Rev.
A{\bf 38}, 1621 (1988).
\item{[6]} S. Nussinov, ``Polarized/Magnetic Versions of
the Casimir Forces", hep-ph/9509228 (September 1995).
\item{[7]} This method, utilizing direct path integration, has been
independently discovered by one of us (O. Keneth (1996),
unpublished), while attempting an exact evaluation of the Casimir
effect for different directions of conductivity suggested in Ref.
(6) above.  We have found that M. Kardar and collaborators (H. Li,
R. Golestanian, M. Goulian) suggested this method sometime ago and
applied it to a broad range of dynamical and other Casimir 
effects--though not to the specific electromagnetic effects of
interest here; see, e.g. H. Li and M. Kardar, Phys. Rev. Lett. {\bf 67},
3275 (1991), Phys. Rev. A{\bf 46}, 6490 (1992), R. Golestanian and
M. Kardar, quant-ph/9802017 (February 1998).
\item{[8]} Since the current $J_\mu$ is conserved, we can adopt this
particular choice.
\item{[9]} Notice that the original one-half factor from Eq.
(10) survives; we have to integrate over both the Real and
Imaginary parts of $J_i(\vec k)$ but thanks to Eq. (9),
only over half the $\vec k$ values say those with
$k_t$ larger than 0.
\item{[10]} It should be emphasized though, that this finite result is
not obtained by discarding what formally looks like ``self-interactions"
pertaining to each plate separately, namely, the $J_1\cdot J_1$ and
$J_2\cdot J_2$ terms.  Indeed, these diagonal
terms are crucial for making the Gaussian integral convergent and
meaningful.
\item{[11]} H. G. B. Casimir and D. Polder, Phys. Rev. {\bf 73}, 369
(1948).  For a more recent, detailed, derivation, see G. Feinberg and
J. Sucher, Phys. Rev. A{\bf 2}, 2395 (1970).
\item{[12]} There are also figure eight-type configurations in which
the B field lines thread the two conducting rings in opposite
directions and these would induce a repulsive interaction.  However,
the B line cannot self-intersect.  The need for spatial avoidance
causes the flux lines to be longer and the action suppression to be
stronger for such configurations.
\end{description}

\section*{Figure Captions}
\begin{description}
\item{Fig. [1]} The Casimir energy versus the angle between
the conducting directions.
\end{description}

\end{document}